\documentstyle[prl,aps,epsf]{revtex}
\begin{document}
\advance\textheight by 0.5in
\advance\topmargin by -0.25in
\draft

\twocolumn[\hsize\textwidth\columnwidth\hsize\csname@twocolumnfalse%
\endcsname

\preprint{NSF--ITP--95--35, cond-mat/9504082}

\title{ {\hfill\normalsize NSF--ITP--95--35, cond-mat/9504082\medskip\\}
Temporal Order in Dirty Driven Periodic Media}

\author{Leon Balents and Matthew P. A. Fisher}
\address{Institute for Theoretical Physics, University of California,
Santa Barbara, CA 93106--4030}

\date{\today}

\maketitle

\begin{abstract}
We consider the non--equilibrium steady states of a driven charge
density wave in the presence of impurities and noise.  In three
dimensions at strong drive, a true dynamical phase transition into a
temporally periodic state with quasi--long--range translational order
is predicted.  In two dimensions, impurity induced phase slips are
argued to destroy the periodic ``moving solid'' phase.  Implications for
narrow band noise measurements and relevance to other driven periodic
media, e.g. vortex lattices, are discussed.
\end{abstract}
\vskip -0.5 truein
]

The influence of quenched impurities on a periodic medium can lead to
very rich physics.  Examples include charge density wave (CDW)
systems\cite{CDW}\ and the mixed state of type II
superconductors\cite{FFH}, in which the vortices form a periodic
lattice.  In both these cases, it has been argued that the impurities
ultimately destroy the long-ranged periodicity and pin the periodic
medium.  However, with an applied force, provided by an electric field
or current, the periodic structure de-pins and becomes mobile.  Once
in motion, the impurities are less effective at destroying the
periodicity.  Indeed, recent experiments on YBaCuO have shown evidence
for a first order melting transition of the {\it moving} vortex
lattice\cite{Vexps}.  Vinokur and Koshelev have interpreted this
experiment in terms of a true non-equilibrium phase transition, from a
moving liquid phase to a ``moving solid phase''\cite{VK}.

This experiment raises a number of questions about the non-equilibrium
steady states of such noisy driven systems with impurities.  The most
basic concerns the very existence of a ``moving solid" phase.  A solid
in equilibrium is usually characterized by the presence of long-ranged
crystalline correlations (Bragg peaks).  But other criteria also
suffice, such as the presence of a non-zero shear modulus, or the
absence of unbound dislocation loops.  Under what circumstances, if
any, is it possible to have a true moving solid, separate from a
driven liquid with plastic flow? If the moving solid phase is
possible, what are it's characteristics and experimental signatures?

In this letter we attempt to answer these questions, focusing for
simplicity on the CDW.  Many of our conclusions, however, apply also
to the driven vortex lattice.  In two dimensions, we find that a
moving solid phase driven through impurities is always unstable to a
proliferation of dislocations.  The system becomes equivalent (in
symmetry) to a driven liquid.  In three dimensions, a moving solid
phase appears to be stable at large velocities, as illustrated in the
schematic phase diagram, Fig.1.  However, the solid phase does not
have true long-ranged positional correlations (LRO), as in an
equilibrium (3d) crystal.  Rather, algebraic power law positional
correlations are predicted, as in a 2d equilibrium crystal.  Likewise,
unbound dislocation loops are absent in the moving solid.  Despite the
absence of spatial LRO, the moving solid phase is periodic in time --
and hence has long-ranged temporal correlations.  The experimental
signature of such a periodic state is narrow band noise, at the
``washboard" frequency\cite{CDW,VLNBN}.

\begin{figure}[hbt]
\epsfxsize=\columnwidth\epsfbox{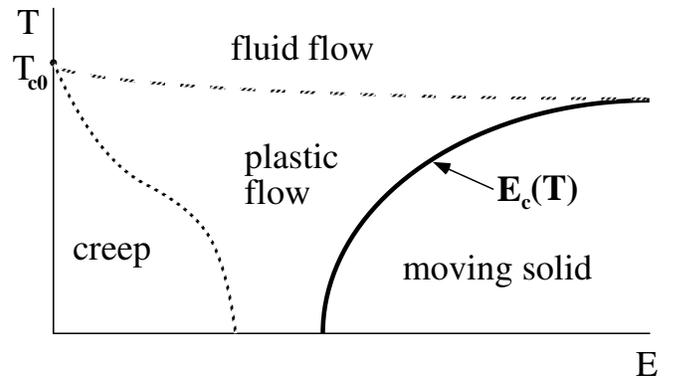}
\vspace{15pt}
\caption{ Schematic phase diagram for the three-dimensional CDW.
Hashed lines indicate (possibly sharp) crossovers.}
\label{fig1}
\end{figure}

Upon inclusion of thermal effects or phase slips, the CDW de-pinning
transition\cite{DSF}\ is predicted to be rounded\cite{Coppersmith},
becoming a crossover (see Fig.1).  For electric fields, $E$, above
this crossover, the CDW is in a plastic flow regime.  With increasing
$E$, we predict a true phase transition into a temporally periodic
``moving solid" phase.  For the CDW, this dynamical transition is
likely to be continuous.  As shown below, scaling arguments then
predict NBN characteristics near the transition.  The phase diagram
for the driven vortex lattice should be similar, with current
replacing electric field.  However, in this case the transition into
the moving solid phase is likely to be first order, at least in the
large current limit.

Charge density waves tend to form in very anisotropic metals,
consisting of weakly coupled metallic chains\cite{CDW}.  The
electronic density in a CDW has a periodic modulation along the chain
(x-)direction: 
\begin{equation}
\rho({\bf x}) = \rho_0 + {\rm Re} \psi e^{i2k_{\rm F}x},
\end{equation}
with $k_{\rm F}$ the in-chain Fermi wave vector.  Long-ranged order of
the CDW is manifest in the complex order parameter field, $\psi({\bf
x})= \rho_1 e^{i\phi}$.

In the absence of impurities and an applied electric field, the CDW
exhibits long-ranged order in the pair correlation function:
\begin{equation}
G({\bf x}) = <\psi^*({\bf x},t)\psi({\bf 0},t)>_t ,
\end{equation}
with $G(x \rightarrow \infty) \ne 0$.  Here the subscript $t$ denotes
a time average (equivalently an ensemble average in equilibrium).  Lee
and Rice have argued that quenched impurities destroy the LRO of $G$,
for physical dimensions, $d<4$\cite{LR}.  However, when the CDW is
driven and moving, the Lee-Rice argument is not valid, and LRO of $G$
is not precluded (but see below).

In the moving non-equilibrium steady state, temporal correlations in
$\psi$ also serve to characterize the CDW order.  Consider the pair
correlation function:
\begin{equation}
C(t) = < \psi^*({\bf x},t)\psi({\bf x},0)>_x ,
\end{equation}
where the subscript $x$ denotes a spatial average.  Temporal LRO is
signaled by a periodic, and non-decaying, behavior: $C(t+t_0)=C(t)$,
with $t_0$ the ``washboard' period.  Middleton has shown that for a
large class of dynamical models, which exclude phase slip and thermal
noise, the steady state is (microscopically) periodic\cite{middleton}.
With noise microscopic periodicity is destroyed, but the
statistical correlation function $C(t)$ is still periodic.  However,
when phase slip is allowed, the robustness of the periodic state is
much less clear, as we discuss below.

To proceed, we assume local CDW order, and construct a long wavelength
description in terms of $\psi$.  In the absence of phase slips, as we
consider first, amplitude fluctuations can be ignored, and the
dynamics involves the phase field, $\phi$.  A common starting point is
the Fukuyama-Lee-Rice (FLR) model\cite{FLR}, with equation of motion:
\begin{equation}
\partial_t \phi = D\nabla^2 \phi + V({\bf x})\sin(2k_{\rm F}x+\phi) +
v2k_{\rm F} .
\end{equation}
The spatial coordinates transverse to the chains have been re-scaled
to give an isotropic diffusion constant $D=\tau v_{\rm F}^2$, with
scattering time $\tau$ and Fermi velocity $v_{\rm F}$.  The second
term on the right side represents the effect of quenched random
impurities.  The last term is present in an applied electric field
$E$, with velocity $v=(e\tau/m) E$.  This term can be shifted away,
$\phi \rightarrow \phi + v2k_{\rm F}t$, reducing the FLR equation to
an equilibrium form: $\partial_t \phi = -\delta H/\delta\phi$, with
Hamiltonian $H$.  However, there are additional terms that can, and
should, be added to FLR, which are manifestly non-equilibrium, as we
now discuss.

The most important such term is $\partial_x \phi$, allowed by symmetry
once the CDW is in motion along the $x$-direction.  The magnitude of
this term can be estimated by balancing forces on a single wavelength,
$\lambda$, of one chain. In an electric field E, the force is $2eE$,
accelerating (locally) the phase field: $2m\partial_t^2 (\lambda
\phi/2\pi) = 2eE$, or equivalently, $\tau \partial_t^2 \phi = v
2\pi/\lambda$.  In the presence of distortions in $\phi(x)$, the
(local) wavelength of the CDW is modified: $2\pi/\lambda \rightarrow
2k_{\rm F} + \partial_x \phi$.  Thus the last term in FLR should be
replaced by: $2k_{\rm F} \rightarrow 2k_{\rm F} + \partial_x \phi$.
(Note that the inertial term has been dropped.)

In general there are other missing terms, for example of the KPZ form,
$(\partial_x \phi)^2$\cite{KPZ}.  This term, involving more gradients
and powers of $\phi$ is less relevant than $\partial_x \phi$.  In the
following we drop this term, although it can play an important
role\cite{BFM}.

In the moving state, it is tempting to argue that the impurity term,
$V(x)$, will average to zero at long times.  However, as pointed out
by Coppersmith\cite{Coppersmith}, it is not legitimate to ignore
completely the effect of impurities.  In particular, the impurities
modify the local mobility, $\mu$, of the phase field.  (This can be
seen explicitly via a high velocity expansion.) The random $V$ term 
can then be replaced by $\delta \mu(x) E 2k_{\rm F} = F(x)$,
where $\delta \mu$ denotes the fluctuating part of the mobility.  We
take $F$ Gaussian with $\overline{F} = 0$ and $\overline{F({\bf
x})F({\bf 0})} = g\delta({\bf x})$.

We thereby arrive at a generalization of FLR:
\begin{equation}
\partial_t \phi = D\nabla^2 \phi + v(2k_{\rm F}+\partial_x\phi) +
F({\bf x}) + \eta({\bf x},t) .
\label{spinwaveeom}
\end{equation}
The stochastic noise term is assumed to be Gaussian with
$\langle\eta\rangle = 0$ and $\langle\eta({\bf x},t)\eta({\bf
x}',t')\rangle= k_B T \delta({\bf x-x'})\delta(t-t')$.

Finally, we modify the model to allow for phase slip processes.  One
way to incorporate phase slip is to put the model on a lattice and
replace $\partial_x \phi \rightarrow a^{-1} sin(\phi(x+a)-\phi(x))$,
etc.  Alternatively, amplitude fluctuations can be included, using
field $\psi$.  The appropriate soft-spin model, which reduces to
Eq.\ref{spinwaveeom}\ in the spin-wave limit, is,
\begin{eqnarray}
\partial_t \psi & = & [D\nabla^2 + v\partial_x + M + r({\bf x}) +
i\omega_0 + iF({\bf x})]\psi \nonumber \\ & & -u\psi |\psi|^2 +
\xi({\bf x},t) ,
\label{softeom}
\end{eqnarray}
with the definition $\omega_0 = v2k_{\rm F}$.  Here $M$ is a ``mass"
term, which controls the magnitude of the order parameter, and
$\xi({\bf x},t)$ is a complex stochastic noise term.  We have also
included a spatially random component to the mass, denoted $r({\bf
x})$, which we take to be a zero--mean Gaussian random variable with
$\overline{r({\bf x})r({\bf 0})} = \Delta \delta({\bf x})$.

First consider the system with phase slips suppressed.  In this
spin--wave limit, Eq.\ref{spinwaveeom}\ is linear in $\phi$ and can be solved via
Fourier transforms:
\begin{equation}
\tilde{\phi}({\bf p},\omega) = {{F({\bf p})} \over {Dp^2 +i v
p_x}}2\pi\delta(\omega) + {{\eta({\bf p},\omega)} \over {i\omega +
Dp^2 + ivp_x}},
\label{phifourier}
\end{equation}
where $\tilde{\phi}(t) = \phi(t) - \omega_0 t$.  The first term in
Eq.\ref{phifourier}\ represents a {\sl static} distortion in $\phi(x)$
induced by the random mobility, while the second gives ``noisy''
dynamical fluctuations around this mean.  The static phase variations
diverge algebraically with system size $L$ for $d < 3$, leading to
(stretched) exponential decay of $G(x)$.  Thus, even without phase
slips, a 2d driven CDW lacks translational LRO.  For the 3d case,
Eq.\ref{phifourier}\ gives $G_{\rm sw}(x) \sim x^{-\eta}$,
corresponding to power law peaks in the static structure function and
translational quasi--LRO (QLRO).  Notice that the presence of the
non-zero $v\partial_x\phi$ term in Eq.\ref{spinwaveeom}\ is critical
here.  Indeed, for a driven periodic system with reflection symmetry,
$v=0$, Eq.\ref{phifourier}\ implies exponential decay of $G(x)$ for
all $d<4$.  Because the disorder term in Eq.\ref{phifourier}\ is
static, the dynamical properties are determined by the thermal noise
term, with $C_{sw}(\omega) \sim \delta(\omega - \omega_0)$ for all
$d>2$, indicating temporal LRO of $C(t)$.  In 2d spin-waves imply
temporal QLRO for $C(t)$.

We now address the stability of these spin-wave results in the
presence of phase slips.  In equilibrium, $v=F=0$, Eq.\ref{softeom}\
describes an XY model with relaxational dynamics.  In this case the
unbinding of {\sl topological} defects (i.e.. vortices) coincides with
the loss of translational LRO due to spin wave fluctuations.  For
$d>2$, the vortices form $d-2$ dimensional subspaces (lines in 3d).
With a core energy growing with size $L$ as $L^{d-2}$, they are bound
at low temperatures.  In equilibrium, 2d is marginal both for spin
waves, which give QLRO, and for vortex unbinding.  But in the
non-equilibrium case of interest, the unbinding of phase slips and
vortices needs to be re-addressed.

For simplicity, consider first the case $v=0$.  Then
Eq.\ref{spinwaveeom}\ can actually
be cast into an equilibrium form, $\partial_t \phi = -\delta E_{\rm
eff}/\delta\phi + \eta$, with the proviso that the ``energy'' $E_{\rm
eff} = \int \{D|\nabla\phi|^2/2 - F\phi\}$ is a multi--valued
(i.e. non--periodic) function of the phase.  The situation is
analogous to the ``tilted washboard'' model of Josephson junctions,
except that the tilt is here a random function of position.

It is clear that spin--wave conformations of the phase are highly
constrained.  Imagine sub--dividing the system into regions of linear
size $L$.  Each such region experiences a net random torque $\int
d^d{\bf r} F({\bf r})$ of order $\pm \sqrt{gL^d}$.  The torque in
neighboring regions is generally different, so that the local phases
are pushed at different rates.  In the absence of phase slips,
however, all regions must rotate synchronously or build up enormous
strains.  Eq.\ref{phifourier}\ describes the resulting steady state in
which the strains increase to counteract the non--uniform applied
torques.

Once phase slips are allowed, however, such a situation is clearly
metastable.  If the net torque in a particular region is positive,
then the energy of the spin--wave state is lowered simply by
increasing all the phases in the region $L$ by $2\pi$.  This decreases
the random energy but does not alter the strain energy (which is now
periodic in $\phi$).  For finite $L$ and non--zero $T$, this process
will therefore occur with an activated rate $1/ \tau \sim
\exp(-U/k_{\rm B}T)$, where $U$ is the energy barrier for the phase
slip process.

The energy $U$ is estimated from the elastic energy mid--way through
the process, i.e. when there exist phase shifts of order $\pi$ on
scale $L$.  Adding the elastic and random contributions to the energy
gives $U(L) \sim D L^{d-2} - \sqrt{gL^d}$.  A more microscopic picture
is that of vortex nucleation.  The phase slip is achieved by
nucleating a small neutral topological defect (vortex--antivortex pair
or vortex loop in $d=2,3$), which expands and slips over the region,
annihilating again on the opposite side.  The elastic contribution to
the barrier energy is just the binding potential of the defect,
$V_{\rm defect}(L) \sim D L^{d-2}$, up to possible $\ln L$ dependence.
For $d >4$ and small $g$, $U(L)$ is positive.  Moreover, since $U(L)$
grows with large $L$ large phase slips are exponentially suppressed,
indicating stability of the spin-wave phase for $d>4$.  For $d<4$,
however, $U(L)$ becomes negative for $L > L_c \sim (D^2/g)^{1/(4-d)}$.
On scales much bigger than $L_c$ it is then inconsistent to assume a
well defined average phase, because of the proliferation of vortex
pairs/loops on smaller scales.  The relaxation time for the phase
slips on scale $L_c$ is $\tau_c \sim \exp[DL_c^{d-2}/k_{\rm B}T]$.
Beyond this time scale, phases separated by distance large compared to
$L_c$ will become de-phased, destroying the temporal LRO of the
spin-wave state.

For $v \neq 0$, the argument is trickier.  First, transform to the
moving frame via $x \rightarrow x - v t$, which removes the
$v\partial_x\phi$ term in Eq.\ref{spinwaveeom}.  The elastic force is
invariant under such a transformation, but $F({\bf r}) \rightarrow
F(x-v t,{\bf r}_\perp)$, so that the random torque field appears to
move with velocity $v$.  Again dividing the system into regions of
size $L$, we see that a statistically uncorrelated realization of $F$
moves into a particular region in a time $t_0 = L/v$.  For large $L$
this decorrelation time $t_0$ is much smaller than the typical
diffusive time for phase changes, $t_{\phi} \sim L^2/D$.  Only on time
scales longer than $t_\phi$ can a phase change take advantage of the
random torques spread out over the entire region.  The random energy
gained is thus averaged over $t_\phi/t_0$ realizations of the torques,
leading to a net torque on the entire region of $F_{\rm net} \sim
\sqrt{g/v}L^{(d-1)/2}$.  In this case the characteristic energy
balance is $U(L) \sim D L^{d-2} -\sqrt{g/v}L^{(d-1)/2}$, and phase
slips proliferate for $d>3$.  Directly in $d=3$, these naive arguments
suggest a transition between an ordered state for $v>v_c$ and a
disordered state for $v<v_c$, with $v_c \sim g/D^2$.

The above arguments are consistent with the spin--wave calculations.
As in equilibrium, they suggest that vortex unbinding coincides with
the loss of translational LRO due to spin-wave variations.  Further
support for this conclusion follows from an analysis of the soft spin
model, Eq.\ref{softeom}, as we now describe.

In the absence of randomness, the soft spin model contains two phases
for $d \ge 2$.  Fluctuation effects near the transition, negligible
above $d=4$, can be studied for small $\epsilon = 4-d$ via the
renormalization group (RG).  The RG including $r({\bf x})$ has been
studied in the context of random--bond XY magnets\cite{Lubensky}; we
generalize this calculation to include $F$ in the dynamics (for
$v=0$).  After transforming $\psi \rightarrow e^{i\omega_0 t}\psi$, we
employ standard dynamical RG methods.  The resulting differential RG
flow equations to quadratic order are,
\begin{eqnarray}
\dot{u} & = & u(\epsilon - 5 u + 6 \Delta), \nonumber \\ \dot{\Delta}
& = & \Delta(\epsilon - 4 u + 4\Delta - 2g) + 2g^2, \nonumber \\
\dot{g} & = & g(\epsilon - 2g + 6 \Delta),
\label{RGflows}
\end{eqnarray}
with $\dot{u} = du/d\ln b$, etc., where $b = e^\ell$ is the rescaling
factor.  A simple analysis shows that the Gaussian ($u=\Delta=g=0$),
pure ($\Delta = g = 0$) and dirty equilibrium ($g=0$) fixed points are
all unstable, and further, that no other fixed points exist.  Instead
the couplings diverge as $\ell \rightarrow \infty$.  In particular, $u
\rightarrow +\infty$ ({\sl not} $-\infty$), so the instability does
not appear to indicate a fluctuation induced first order transition.
Instead, the strong divergence of the disorder strengths $g$ and
$\Delta$ are consistent with the scenario that the ordered phase is
absent.

For $v \neq 0$, changing to co--moving coordinates $x \rightarrow x -v
t$ reduces Eq.\ref{softeom}\ to the previous case but with $r({\bf x})
\rightarrow r(x-vt,{\bf x}_\perp)$ and $F({\bf x}) \rightarrow
F(x-vt,{\bf x}_\perp)$.  Because $z=2+O(\epsilon)$ at the pure XY
fixed point, the weaker $x$ dependence of $r$ and $F$ may be ignored.
Perturbations of the form $r(-vt,{\bf x}_\perp) \psi$ and $iF(-vt,{\bf
x}_\perp)\psi$ are strongly irrelevant near $d=4$, consistent with the
spin-wave analysis and our earlier scaling arguments which gave d=3 as
the lower critical dimension for the ordered phase.  To study d=3 in
the soft spin representation, we employ non-perturbative techniques.
We therefore consider a generalized model containing $N$ complex
fields $\psi_i$ obeying Eq.\ref{softeom}, but with $|\psi|^2
\rightarrow \sum_i |\psi_i|^2$.  We analyze the stability of the pure
$N=\infty$ fixed point to the random perturbations.  From scaling the
singular part of the mean energy density varies as $\overline{\langle
\psi^2 \rangle} = \xi^{1/\nu - d}f(\Delta \xi^{y_\Delta}, g
\xi^{y_g})$, where $\xi \sim M^{-1/\nu}$ is the (pure) correlation
length, and $y_\Delta$ and $y_g$ are the RG eigenvalues of $\Delta$
and $g$, respectively.  At $N=\infty$, $1/\nu -d = -2$.
Differentiation implies $A_\Delta \equiv \partial_\Delta
\overline{\langle \psi^2 \rangle}|_{\Delta,g=0} \sim
\xi^{y_\Delta-2}$, $A_g \equiv \partial_g \overline{\langle \psi^2
\rangle}|_{\Delta,g=0} \sim \xi^{y_g-2}$.  These quantities are
computed at $N=\infty$ using saddle point techniques and the
Martin--Siggia--Rose dynamical formalism\cite{MSR}.  We find
$y_\Delta = d-5$ and $y_g = 3-d$.  (As a check we also considered the
case $v=0$, and found $y_\Delta = d-4$ and $y_g = 4-d$, in agreement
with the usual Harris criterion and the
$\epsilon$-expansion, Eq.\ref{RGflows}.)  Thus for $d<3$, the equilibrium
critical point is unstable to random $F$, consistent with the absence
of an ordered phase.

Our predictions for the 3d phase diagram are summarized in Fig.1.
Upon lowering the temperature at weak drive, $E$, substantial CDW
amplitude develops at the mean-field transition temperature $T_{c0}$.
At long distances and times, however, both $G(x)$ and $C(t)$ decay
exponentially to zero.  With increasing drive, the CDW undergoes a
sharp non-equilibrium phase transition into an ordered ``periodic
state" with spatial QLRO and temporal LRO.  Our arguments strongly
suggest that for 2d CDW systems, the ordered phase is absent.

Experimentally, temporal LRO in the solid phase manifests
itself in NBN.  Consider current fluctuations in the presence of a
fixed bias voltage; other set--ups are qualitatively similar.  The
current density $j({\bf x}) = (e n_\perp/\pi) \partial_t\phi[1 +
(\rho_1/2k_{\rm F})\cos(2k_{\rm F}x + \phi)]$, where $n_\perp$ is the
areal chain density.  In a sample of cross--sectional area $A$, the
instantaneous CDW current through the plane $x=0$ is
$I_\times(t) = \int_A \! d^2{\bf x}_\perp j(x=0,{\bf x}_\perp,t)$.
The oscillatory part of the NBN correlator, $S(t) \equiv \langle
I_\times(t)I_\times(0)\rangle$ is thus
\begin{equation}
S_o(t) \approx I_0^2 \int_{\bf x_\perp,x_\perp'}\!\!\!\!\!\!{\rm
Re}\{e^{i\omega_0 t}\left\langle \psi({\bf x_\perp},t)\psi^*({\bf
x_\perp'},0)\right\rangle\},
\label{currentflucts}
\end{equation}
where $I_0 = (en_\perp\omega_0)/(2 \sqrt{2} k_{\rm F}\pi)$.  We
consider this quantity in the bulk, and expect that measured current
fluctuations (in the external leads) exhibit proportional behavior.
Temporal LRO in the solid phase therefore implies a sharp (resolution
limited) delta function peak in $S(\omega) \sim A^{2-\eta/2}\delta
(\omega - \omega_0)$.  Deep in the liquid phase, $\psi$ correlations
are short--range in space and time, which gives the mean--field result
$S(\omega) \sim A \Omega /(\Omega^2 + (\omega-\omega_0)^2)$.  Near the
transition field $E_c(T)$, provided the transition is continuous, we
expect a scaling form $S_\pm(\omega,\delta E) \sim |\delta E|^a
f_\pm[(\omega-\omega_0)/|\delta E|^{z\nu},A|\delta E|^{2\nu}]$, where
$z$ and $\nu$ are the dynamical and correlation length exponents, $a$
is an additional scaling exponent, and $\delta E=E-E_c$.  Matching to
the infinite area limit implies that the amplitude of the delta
function frequency peak for $E>E_c$ vanishes as $|\delta
E|^{a+(4+z-\eta)\nu}$.  For $E<E_c$, matching implies the (generally
non--Lorentzian) line-shape $S(\omega) \sim A|\delta E|^{a+2\nu}
s[(\omega-\omega_0)/|\delta E|^{z\nu}]$.  In two dimensions,
$S(\omega)$ has an intrinsic width for all fields and temperatures.

Although the discussion has focused on CDWs, most of the ideas
employed here apply to more general periodic media.  Of particular
current experimental interest are vortex lattices and 2d Wigner
crystals\cite{Wigner}. In all cases, translational and temporal LRO
may be destabilized both by phonons (phase fluctuations) and by
topological defects (phase slips).  Provided reflection invariance is
broken by an external drive field, we expect linear gradient terms
(e.g. $v\partial_x\phi$) in the equations of motion.  Preliminary
investigation of driven lattices suggests that such terms play a
similar role in that case\cite{BFM}.  There are also examples of
oscillatory states in pattern forming systems without broken
reflection symmetry -- leading, we expect, to the absence of the
ordered phase in the presence of disorder and noise.

We are grateful to J. Krug, D.S. Fisher, and G. Grinstein for helpful
conversations.  This work has been supported by the National Science
Foundation under grants No. PHY94--07194 and No. DMR--9400142.

\end{document}